\documentclass[usenatbib]{mnras}

\usepackage{multirow}
\usepackage{amsmath}
\usepackage{amssymb}
\usepackage{graphicx}	

\def\citetJuric{J08}
\long\def\Ignore#1{\relax}

\def\tabpar#1{\hsize=13cm\parindent=0pt\baselineskip=11pt\raggedright\vtop{#1}}
\def\tabcol#1{{\hsize=3cm\parindent=0pt\baselineskip=11pt\raggedright\vtop{#1}}}

\newcommand{\Md}        {M_{\rm d}}
\newcommand{\Rc}        {R_{\rm c}}
\newcommand{\Rd}        {R_{\rm d}}
\newcommand{\zd}        {z_{\rm d}}

\newcommand{\Vc}        {v_{\rm c}}
\newcommand{\fDM}        {f_{\rm DM}}
\newcommand{\fP}        {f_{\rm P}}

\newcommand{\kms}{\,{\rm km\, s^{-1}}}
\newcommand{\Msun}{\,{\rm M}_\odot}

\newcommand{\df} {DF}
\newcommand{\HI} {{H{\sc{i}}}}
\newcommand{\vJ} {\mathrm{\textbf{J}}}
\newcommand{\vL} {\mathrm{\textbf{L}}}
\newcommand{\vr} {\mathrm{\textbf{r}}}
\newcommand{\kpc} {\,\mathrm{kpc}}
\newcommand{\pc} {\,\mathrm{pc}}
\newcommand{\cm} {\,\mathrm{cm}}
\newcommand{\GeVc}{\,\mathrm{GeV/c^2}}
\newcommand{\e} {\mathrm{e}}
\newcommand{\rd} {\mathrm{d}}
\newcommand{\msun}{\Msun}
\def\Gyr{\,{\rm Gyr}}
\newcommand{\cut}[1]{}

\title[A centrally heated dark halo for our Galaxy]{A centrally heated dark
  halo for our Galaxy} \author[D. R. Cole and James Binney]{D.R.Cole$^{1}$\thanks{E-mail:
    {\tt david.cole@physics.ox.ac.uk}} and James Binney$^{1}$\thanks{E-mail:
    {\tt binney@physics.ox.ac.uk}} \\
$^{1}$Rudolf Peierls Centre for Theoretical Physics, Keble Road, Oxford, OX1 3NP, United Kingdom}

\begin{document}

\date{Accepted xxx Received xxx ; in original form \today}

\maketitle

\label{firstpage}


\begin{abstract}
We construct a new family of models of our Galaxy in which dark matter and
disc stars are both represented by distribution functions that are analytic
functions of the action integrals of motion.  The potential that is
self-consistently generated by the dark matter, stars and gas is determined,
and parameters in the distribution functions are adjusted until the model is
compatible with observational constraints on the circular-speed curve, the
vertical density profile of the stellar disc near the Sun, the kinematics of
nearly 200$\,$000 giant stars within $2\kpc$ of the Sun, and estimates of the
optical depth to microlensing of bulge stars. We find that the data require a
dark halo in which the phase-space density is approximately constant for
actions $|\vJ|\lesssim140\kpc\kms$. In real space these haloes have core
radii $\simeq2\kpc$.
\end{abstract}



\section{Introduction}  
\label{sec:intro}

It is now generally accepted that most of the mass of galaxies like ours is
contained in a dark halo that is made up of particles that have yet to be
discovered. Since these dark-matter (DM) particles have so far only come to
our notice through the gravitational field that they generate, the only way
to discover how they are distributed is to model that gravitational field
using tracer particles. The tracer particles range from photons (via
gravitational lensing), through electrons (via thermal X-ray emission) to
stars and neutral atoms. Stars and atoms are the tracer particles of choice
in our own Galaxy.

We understand the dynamics of our Galaxy at and inwards of the solar radius
$R_0$ much better than we do at large radii, in part because determining
distances to and tangential velocities of tracers at $R>R_0$ is hard, and in
part because the density of stars and neutral gas is small at $R\gg R_0$, so
statistical uncertainties become large.

One constrains the distribution of DM by building dynamical models of
our Galaxy that are consistent with relevant data. These data include
(i) the circular-speed curve $\Vc(R)$ that one extracts from
radio-frequency emission lines of interstellar gas, the proper motion of Sgr
A* and the kinematics of stellar masers; (ii) the
kinematics of stars that lie close enough to the Sun to have useful
proper motions; (iii) star counts, which strongly constrain the
vertical structure of the stellar disc at $R\sim R_0$; (iv)
measurements of the optical depth to microlensing of bulge stars,
since these measurements constrain the amount of stellar as opposed to
interstellar or dark mass at $R\la6\kpc$.

As a first approximation one usually assumes that the Galaxy is axisymmetric.
Then equilibrium dynamical models of the stars and DM are most conveniently
formulated in terms of a distribution function (DF) $f(\vJ)$ that depends on
the three action integrals $J_r$, $J_\phi\equiv L_z$ and $J_z$. $J_r$
quantifies a star's radial excursions, $J_z$ quantifies its oscillations
perpendicular to the Galactic plane, and $J_\phi$ is the component of angular
momentum $\vL$ abut the symmetry axis.

\cite{Piffl2014} (hereafter P14) used models of this type to obtain very tight limits on the
mass of DM at $R\le R_0$. The central idea of this work is that the
kinematics of stars in the RAdial Velocity Experiment
\citep[RAVE][]{Steinmetz2006} essentially fix the dependence of $f$ on $J_r$ and
$J_z$, and this dependence essentially fixes the run of vertical
gravitational force $K_z(z)$ required to produce the stellar density profile
$\rho_*(R_0,z)$ determined from star counts. The required $K_z(z)$ is
produced by a combination of matter in the disc and the dark halo, and if
the shape of each contribution is known, both normalisations can be
recovered from $K_z(z)$. Thus the data strongly constrain the local surface
density of stars and the volume density of DM. If one now assumes, as
P14 and all earlier authors did, that one knows the functional
form of the DM density profile, then the entire structure of the dark halo
can be fixed from local data. The scale-length of the disc then follows from
the measured circular-speed curve $\Vc(R)$. 

The major uncertainty in this modelling is the flattening of the dark halo.
P14 found a one-parameter family of successful models with dark
haloes that had essential the same mass inside the isodensity surface through
the Sun but differed in the axis ratio $q$ of those surfaces. The flatter a
dark halo was, the more it contributed to both $\Vc(R_0)$ and $K_z$, and
therefore the smaller was the required mass of the disc.

P14 assumed that in the spherical case the density profile of the dark halo
is the NFW profile \citep{Navarro1997}
\begin{equation}
\label{eq:NFW}
\rho(r)=\frac{\rho_0}{r/r_{\rm s}(1+r/r_{\rm s})^2},
\end{equation}
 where $\rho_0$ and $r_{\rm s}$ are constants.  This profile fits dark haloes
in cosmological simulations that do not contain baryons.  \cite{Piffl2015}
(hereafter P15) observed that we would not expect this profile to fit the
Galaxy's dark halo, since the latter is subject to the disc's non-trivial
gravitational field.  Since \cite{Blumenthal1984} it has been argued that in
galaxies like ours, in which baryons have accumulated steadily over a long
period, the dark halo would have responded to the strengthening gravitational
field adiabatically -- for recent work see \cite{Katz2014}.  That is, the DF
of the halo particles $\fDM(\vJ)$ would be invariant as the disc and bulge
accumulated. Consequently, P15 replaced the assumption of an NFW density
profile with the assumption that $\fDM(\vJ)$ had a certain form, which they
chose such that, in the absence of a disc or bulge, the dark halo would
coincide with the NFW halo fitted to the data by P14. When the DF of the disc
was also set to that found by P14, the resulting model violated the
constraints on $\Vc(R)$ at small $R$. Thus, when near the Sun the balance
between disc and halo mass is that required by the data and one takes into
account the tendency of the dark halo to deform elastically as the disc and
bulge grow, excessive mass accumulates near the Galactic centre.

The process of searching model space for models that are consistent with the
data becomes much more costly computationally when the dark halo is specified
by $\fDM(\vJ)$ rather than $\rho(\vr)$. Hence P15 only computed a single
model, that was based on the parameters found by P14.  \cite{Binney2015}
(hereafter BP15) conducted a systematic search for a model that (i) has a
dark halo specified by a DF $\fDM(\vJ)$ that would in isolation generate an
NFW profile, and (ii) is consistent with $\Vc(R)$ and the local kinematic and
star-count data. They found such a model. It avoided placing too much mass at
small radii by assigning the disc a large scale radius $\Rd$. As a
consequence, DM dominated the gravitational force down to small radii, and in
this case the optical depth to lensing bulge stars falls below observational
requirements. The clear conclusion from this exercise is that the DM
did {\it not} respond adiabatically to the accumulation of baryons, so now
$\fDM(\vJ)$ would not in isolation generate an NFW profile.  This conclusion
is actually not unexpected because in the NFW model the phase-space density
tends to infinity at the origin of action space. While it is in principle
possible that the infinite phase-space density of the CDM initial conditions
survives structure formation at the centres of dark haloes, it is improbable
that scattering of DM particles has bot flattened the DM
density near the centres of haloes.

The purpose of this paper is to obtain a form of $\fDM(\vJ)$ that {\it is}
consistent with all the observational data and our understanding of
cosmology. In Section~\ref{sec:observ} we summarise the observational
constraints imposed on models. In Section~\ref{sec:DMDF} we modify a DF that
generates an NFW halo so the phase-space density of DM tends to a
constant for the most bound particles. In Sections~\ref{sec:discDF} to
\ref{sec:fitting} we specify the other components of our model Galaxy, which
include DFs for the thin and thick discs, and explain how the parameters of
the DFs are adjusted to obtain self-consistent models that are compatible
with the observational constraints. Section~\ref{sec:results} describes the
some successful models. Section \ref{sec:mulens} examines these models from
the perspective of the microlensing data. Section~\ref{sec:nogood} we
investigate the symptoms of adopting an excessively large core for the dark
halo. In Section~\ref{sec:discuss} we review studies of the formation of
cored dark halos.  In Section~\ref{sec:conclude} we sum up and look to the
future.

\section{Observational inputs}  
\label{sec:observ}

We used the same observational inputs as BP15. Here we summarise the inputs;
more detail can be found in BP15. Our solar
parameters are given in Table \ref{tab:sol}.

\begin{table}
  \caption[]{Adopted position and velocity of the Sun.}
\label{tab:sol}
  \begin{center}
    \begin{tabular}{lcc} 
      \hline
      \multicolumn{1}{l}{Parameter} &
      \multicolumn{1}{c}{} &
      \multicolumn{1}{c}{source}
 \\[-0.5ex]
      \hline
\hline
 $R_0/\kpc$ & 8.3 & \cite{Schoenrich2012} \\
 $z_0/\kpc$ & 0.014 & \cite{Binney1997} \\
 $\textbf{V}_\odot/\kms$ & (11.1,12.24,7.25) &\cite{Schoenrich2010}\\
\hline
    \end{tabular}
  \end{center}
\end{table}

\subsection{Gas terminal velocities}
The distribution of \HI\ and CO emission in the
longitude-velocity plane yield a characteristic maximum (``terminal'')
velocity for each line of sight
\citep[e.g.][\S9.1.1]{Binney1998} which are
related to the circular speed $\Vc(R)$. We use the terminal
velocities $v_{\rm term}(l)$ from \citet{Malhotra1995}.  Following
\citet{Dehnen1998} and \citet{McMillan2011} we neglect data at $|\sin l|
< 0.5$ in order not to be influenced by the Galactic bar, and we
assume that the ISM has a Gaussian velocity distribution of dispersion
7~km\,s$^{-1}$.
\subsection{Maser observations}
We use 103 maser observations from \citet{Reid2014} that provide
6D phase-space information. The maser sources, which are associated with
young stars, are assumed to be on nearly circular orbits:
their velocities are assumed to be Gaussianly distributed about the circular
velocity with  dispersion $7\kms$
\citep{vanderKruit1984,McMillan2010}. For the likelihood computation
we neglected all sources at $R < 4\kpc$ to prevent the Galactic bar giving
rise to  a bias.
\subsection{Proper motion of SgrA*}
We adopt from   \cite{Reid2004} the proper motion
\begin{equation}
 \mu_\mathrm{SgrA^\star} = -6.379 \pm 0.024~\mathrm{mas\,yr}^{-1}.
\end{equation}
 of the radio source SgrA* associated with the super-massive black hole that
sits in the Galactic Centre, as an estimate for the solar motion with
respect to the GC.
\subsection{Vertical density profile from SDSS}

We assume that the population from which the RAVE sample is drawn is
identical to that studied by \citet{Juric2008} (hereafter \citetJuric). We
use the data points shown in the middle panel of Figure~15 in \citetJuric,
which shows results from M dwarf stars in the colour range $0.70 < r-i <
0.80$, this sample should carry only weak biases in metallicity and age.  We
omitted the correction of the data for the effects of Malmquist bias and
binarity as they had a negligible effect on the results of P14 and we
decomposed the density profile into contributions from the disc and stellar
halo as in BP15.

\subsection{Kinematics from RAVE}

We use the stellar parameters and distance estimates in the fourth
RAVE data release \citep{Kordopatis2013}. We sort the stars into eight
spatial bins, four inside the solar cylinder and four outside. We
compute the velocity distributions predicted by our \df\ at the mean
$(R,z)$ positions (barycentre) of the stars in each bin. We
have a histogram for each velocity component, so we accumulate $\chi^2$
from these 24 histograms.

We modify the model distributions to take into account the effect of
errors in the velocity and parallax estimates in the data. This
procedure is fully described in P14. Our model
selection involves optimising the fit between the data and the
velocity histograms after the latter have been modified to allow for
the impact of errors in the measurements of velocity and distance.

\section{Modelling procedure}  
\label{sec:model}

Most of the components used in our model are the same as those used
by BP15, the main difference being the dark halo. We
describe our dark halo \df\ first and then summarise the other components.

\subsection{Distribution Functions for heated DM}  
\label{sec:DMDF}

The central density cusp of the NFW model implies divergence of the
phase-space density $f$ of DM particles as their action integrals $J_i$ go to
zero because in the cusp the velocity dispersion must tend to zero.
Quantitatively, \cite{Posti2015} showed that the \df\
\begin{equation}
\label{eq:Posti}
\fP(\vJ)=\frac{N}{J_0^3}\frac{(J_0/h)^{5/3}}{(1+h/J_0)^{2.9}}
\end{equation}
with $h(\vJ)$ a homogeneous function of degree unity,
self-consistently generates a system that closely resembles the NFW profile,
with the scale action $J_0$ encoding  the scale radius around which the
slope of the radial density profile shifts from $-1$ at small radii to $-3$
far out.
From equation (\ref{eq:Posti}) it follows that
$f_P(\vJ)\sim |\vJ|^{-5/3}$ as $|\vJ|\rightarrow0$.

The message from BP15 is that the \df\ of DM cannot increase as strongly as
$|\vJ|\rightarrow0$ as an NFW profile predicts, either because the infinite
phase-space density of the CDM initial conditions does not survive structure
formation even at the centres of dark haloes, or because the baryons did not
accumulate entirely adiabatically, and the fluctuating gravitational
potential associated with them has upscattered what were the most tightly
bound DM particles.  This conclusion parallels what has been learnt from
studies of dwarf galaxies: their rotation curves rise less steeply near their
centres that would be expected if their dark haloes had cental density cusps
like that of the NFW profile \citep[][and references therein]{AgnelloAE},
again implying that at some stage even the most bound DM particles have been
upscattered.

We seek to model only upscattering of the most bound DM particles --
away from the centre of a dark halo the DM particles have been abundantly
scattered to low phase-space densities by the potential fluctuations
associated with structure formation. We need to model scattering, by whatever
agent, that is not correctly captured by the NFW model.
This line of argument motivates us to propose analytic approximations to the
current \df\ of Galactic DM based on the assumption that in action space
upscattering has produced a constant-density core, from which the DM density
falls monotonically as $|\vJ|$ increases such that it asymptotes to the
DM-only form (\ref{eq:Posti}) proposed by \cite{Posti2015}.

To obtain a satisfactory DF we multiply the \df\ (\ref{eq:Posti}) by a
function $g(h)$ that varies as $h^{5/3}$ for small $h$ and tends to unity for
large $h$. At intermediate values of $h$, $g$ should exceed unity by a small
amount to ensure that the total mass of DM is conserved. That is, we require
\begin{equation}
\label{eq:intcond}
\int\rd^3\vJ\,f(h)=\int\rd^3\vJ\,g(h)\fP(h).
\end{equation}

\begin{figure}
\includegraphics[width=\hsize]{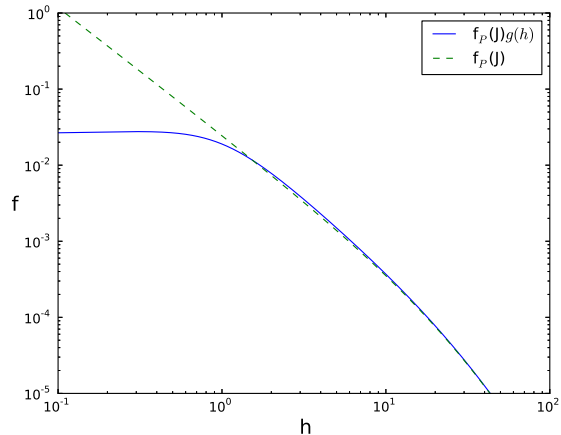}
\caption{Dashed curve: the \citep{Posti2015} \df\ (eqn \ref{eq:Posti})  with
$J_0=2000\kpc\kms$. Full curve:
the \df\ obtained by multiplying by the function $g(h)$ with
$h_0=100\kpc\kms$.}\label{fig:defsg}
\end{figure}
A suitable functional form for $g$ is
\begin{equation}
\label{eq:defsg}
g(h)=\left[{h_0^2\over h^2}-\beta{h_0\over h}+1\right]^{-5/6},
\end{equation}
 where $h_0$ is an arbitrary constant with the dimensions of action that sets
the scale of the almost constant-density core of the final \df\ $g\fP$, and
the constant $\beta$, which controls the magnitude in the peak of $g$ before
it asymptotes to unity, is determined by requiring satisfaction of equation
(\ref{eq:intcond}).  

The function $g$ is promising for present purposes,
because (i) when $h\ll h_0$, the denominator is dominated by the first term,
so $g(h)\simeq(h/h_0)^{5/3}$ which is what is required to annul the
divergence of $\fP$ as $h\to0$, and (ii) when $h\gg h_0$ it is evident that
$g\to1$ as required. The peak of $g$ is associated with the minimum of the
quantity in square brackets, which occurs when $h=2h_0/\beta$ and
$g=1/(1-\beta^2/4)$. Naturally we require $\beta<2$ to ensure that the
quadratic expression on the bottom of equation (\ref{eq:defsg}) has no real
roots. The full curve in Fig.~\ref{fig:defsg} shows the \df\ obtained for
$J_0\equiv2000\kpc\kms$ and $h_0=100\kpc\kms$.

The NFW halo extends to infinity, but we require a halo in which the density
vanishes sufficiently far from the origin to simplify the computation of the
the potential generated by the halo. Since we are exclusively interested in
the structure of the halo at radii $r\la20\kpc$, there is no reason not to
truncate the halo at a large radius. This we do by subtracting from $g\fP$
with argument $h$ the value of $g\fP$ evaluated at $h_{\rm max}$ and
declaring the DF to be zero if $h>h_{\rm max}$. That is our final \df\ is
\begin{equation}
\fDM(\vJ)=
\begin{cases}g(h)\fP(h)-g(h_{\rm max})\fP(h_{\rm max})&\hbox{for }h<h_{\rm
max}.\\
0&\hbox{otherwise.}
\end{cases}
\end{equation}
The constant $h_{\rm max}$ can be any a large action that prevents the dark
halo extending at low density to infinity. Our choice, $h_0=10^6\kpc\kms$,
has negligible effect on the halo's density within the radius
$R_{200}\sim250\kpc$ at which the halo's density becomes 200 times the mean
cosmic density of matter. Consequently, no reported property of the halo
would be changed by increasing $h_{\rm max}$ to arbitrarily large values.

The  homogeneous function $h$ controls the flattening and velocity anisotropy
of the halo. Following BP15 we adopt
\begin{equation} \label{eq:hJ}
 h(\vJ)= {1\over A}J_r + \frac{\Omega_\phi}{B\kappa} (|J_\phi| +
 J_z),
\end{equation}
 where $A$ and $B$ are given by equations (6) and (7) of P15 with $b=8$ to
ensure radial anisotropy. Appendix C of P15 gives the rationale for this
choice of the dark halo's \df. On account of the appearance of ratios of
epicycle frequencies in equation \eqref{eq:hJ}, the functional dependence on
$\vJ$ of our halo DF depends on the model's potential. When the disc and
bulge are introduced, this circumstance is inconvenient, so after we have
determined the potential that a halo DF generates in isolation, we freeze the
functional forms $\Omega(J_\phi)$, $\kappa(J_\phi)$ and $\nu(J_\phi)$ of the
frequencies that appear in the definition of $h(\vJ)$. Consequently, in the
final model the frequencies of circular orbits are only approximately given
by these functions.

Since $\beta$ is determined by equation (\ref{eq:intcond}), the free
parameters in $\fDM$ are $J_0$, which sets the NFW scale radius, $h_0$, which
sets the size of the dark halo's core, and the normalisation $N$.

\begin{figure}
\includegraphics[width=1.1\hsize,angle=0]{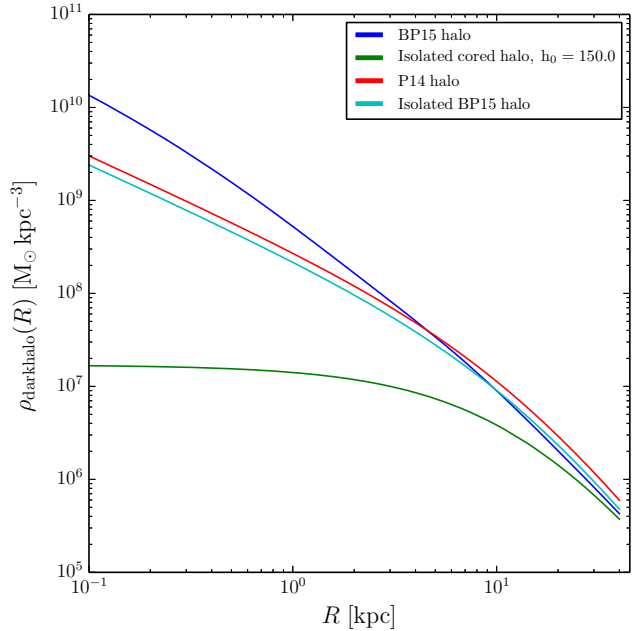}
\caption{Density profiles of some dark haloes. The dark green curve shows the
profile of an isolated cored DM halo with $h_0=150\kpc\kms$. The red curve is
for the NFW profile from P14. The cyan and  blue curves show the profiles
generated by the BP15 DF in isolation and in the
presence of the baryons, respectively.}
\label{fig:DMhalo}
\end{figure}

The dark green curve in Fig.~\ref{fig:DMhalo} shows the density profile that
is generated in the plane $z=0$ by the halo DF of one of our final models
when the halo is isolated. Clearly this halo has a homogeneous core that
extends to $\sim2\kpc$. We show for comparison the density profiles of:
(i) in red the NFW halo flattened to axis ration $q=0.8$ that P14 fitted;
(ii) in dark blue the dark halo fitted by BP15; (iii) in magenta the halo
that the BP15 halo DF generates in isolation.  The difference between the
dark blue and magenta curves represents the adiabatic deformation of the dark
halo by the disc and bulge. The red curve has the same form as the magenta
curve because the BP15 DF was one which in isolation generates an NFW
profile, and the red curve lies above the magenta curve because the former is
a fit to the density of the halo in the presence of the disc rather than in
isolation. The dark green curve lies below the other curves because it shows
what the current halo would look like if the disc and bulge were to be slowly
dismantled, allowing the halo to expand.  Hence it is most comparable to the
magenta curve. It lies below this curve because our dark halo is less massive
than that of BP15, and thus allows the disc to place more mass at $r<R_0$, and
in this way provide adequate optical depth for microlensing.

\subsection{The stellar disc}\label{sec:discDF}

As in P14 and BP15, the DF of the disc is superposition of the
``quasi-isothermal'' form that was introduced by \cite{Binney2011}, namely
\begin{equation}
\label{eq:qi}
 f(J_r,J_z,J_\phi)=f_{\sigma_r}(J_r,J_\phi)f_{\sigma_z}(J_z,J_\phi),
\end{equation}
where $f_{\sigma_r}$ and $f_{\sigma_z}$ are
\begin{equation}
\label{planeDF}
 f_{\sigma_r}(J_r,J_\phi)\equiv \frac{\Omega\Sigma}{\pi\sigma_r^2\kappa}
 [1+\tanh(J_\phi/L_0)]\e^{-\kappa J_r/\sigma_r^2}
\end{equation}
and
\begin{equation}
\label{basicvert}
 f_{\sigma_z}(J_z,J_\phi)\equiv\frac{\nu}{2\pi\sigma_z^2}\,
 \e^{-\nu J_z/\sigma_z^2}.
\end{equation}
Here $\sigma_r(J_\phi)$ and $\sigma_z(J_\phi)$ control the radial and
vertical velocity dispersions, while $\Omega(J_\phi)$, $\kappa(J_\phi)$ and
$\nu(J_\phi)$ are, respectively, the circular, radial and vertical epicycle
frequencies of the circular orbit with angular momentum $J_\phi$. The function
\begin{equation}
\label{eq:defsSigma}
 \Sigma(J_\phi)=\Sigma_0\e^{-\Rc/\Rd},
\end{equation}
where $\Rc(J_\phi)$ is the radius of the circular orbit, determines the surface
density contributed by the quasi-isothermal. To keep the disc scale-height
roughly independent of $R$, we take $\sigma_z\sim\exp(-\Rc/R_\sigma)$, where
the constant $R_\sigma\sim2\Rd$. Although there is no compelling reason to do
so, it is customary to give $\sigma_r$ the same dependence on $J_\phi$.

The DF of the thick disc is taken to be a single quasi-isothermal, while the
thin disc's DF is built up out of a quasi-isothermal for each coeval cohort
of stars.

The \df\ of the thin disc is taken to be a superposition of quasi-isothermal
\df s, one for the stars of each age $\tau$, and the  velocity-dispersion
parameters $\sigma_i$ depend on the age $\tau$ of the cohort in addition to
$J_\phi$. As in earlier papers we assume
\begin{equation}
\sigma_0(\tau)= \left(\frac{\tau+\tau_1}{\tau_{\rm T}+\tau_1}\right)^\beta\sigma_0.
\end{equation}
 where $\beta=0.33$, $\tau_1=0.11\Gyr$ is a constant that determines the velocity dispersion of
stars at birth, $\tau_{\rm T}=10\Gyr$ is the present age of the disc, and
$\sigma_0$ is the current velocity dispersion of the oldest thin-disc
stars near the Sun.

We assume that the star-formation rate in
the thin disc has decreased exponentially with time, with characteristic time
scale $t_0=8\Gyr$, so the complete thin-disc \df\ is
\begin{equation}\label{thinDF}
 f_{\rm thn}(J_r,J_z,J_\phi)=\frac{\int_0^{\tau_{\rm m}}\rd\tau\,\e^{\tau/t_0}
 f_{\sigma_r}(J_r,J_\phi)f_{\sigma_z}(J_z,J_\phi)}{t_0(\e^{\tau_{\rm m}/t_0}-1)}.
\end{equation}
 We set the normalising constant $\Sigma_0$ that appears in
equation (\ref{eq:defsSigma}) to be the same for both discs and use for the
complete \df
\begin{equation}
 f_{\rm disc}(J_r,J_z,J_\phi)=f_{\rm thn}(J_r,J_z,J_\phi) +
 F_{\rm thk}f_{\rm thk}(J_r,J_z,J_\phi),
\end{equation}
where $F_{\rm thk}$ is a parameter that controls the fraction $(1+F_{\rm
thk}^{-1})^{-1}$ of stars that belong to the thick disc. The values of the
parameters for our final model are given in Table~\ref{tab:discs}.

We followed P14 in imposing a lower limit of $1\kpc$ on
the value of $\Rc(J_\phi)$ at which the epicycle frequencies
$\kappa(J_\phi)$ and $\nu(J_\phi)$ are evaluated for use in the \df.

\subsection{DF of the stellar halo}\label{sec:starhalo}

As in P14 and BP15, when computing velocity histograms for comparison with
the RAVE data, we add to the stellar DF  a small contribution from the
stellar halo, which we presume to have no net rotation. If a halo population
is not included, the DF of the thick disc is distorted to provide some
stars that are counter-rotating to the disc, since the data include a few
such stars. Since the mass of the stellar halo is negligible, we do not
include the DF of the stellar halo when we integrate over velocities to
determine the total stellar mass. Since we only require the stellar halo at
points near the Sun, it is adequate to adopt from \cite{Posti2015} the DF
that generates a power-law density profile  $\rho_{\rm halo}
\propto r^{-\alpha}$, with index $\alpha \simeq 3.5$ \citep[see
e.g.][\S10.5.2]{Binney1998}. Thus we take as the (un-normalised) \df\ of the
stellar halo to be
\begin{equation}\label{eq:haloDF1}
f(\vJ)  = h_*^{3.5}\exp\{-[h_*/h_{*\rm max}]^4\},
\end{equation}
where
\begin{equation}
h_*(\vJ)\equiv J_r + \gamma_1|J_\phi| + \gamma_2 J_z + J^{\rm(s)}_{\rm core}.
\end{equation}
Here $\gamma_1 = 0.937$, $\gamma_2 = 0.682$, $J^{\rm(s)}_{\rm core} =
200\kpc\kms$ and $h_{*\rm max} = 2.5\times10^5\kpc\kms$. With these choices
the 
stellar halo is approximately spherical. 

In summary, when computing kinematics, the  total stellar \df\ is taken to be
\begin{equation}
 f(J_r,J_z,J_\phi) = f_{\rm disc}(J_r,J_z,J_\phi) + F_{\rm halo}f_{\rm halo}(J_r,J_z,J_\phi)
\end{equation}
with $F_{\rm halo}$ chosen so $\rho_{\rm halo}/\rho_{\rm disc}=0.0056$
at $(R,|z|)=(R_0,0.5\kpc)$ to be consistent with the \citetJuric\ data as
explained in BP15.

\subsection{The bulge/bar and gas disc}\label{sec:bulgegas}

Our modelling technique restricts us to axisymmetric models, so we cannot use
a sophisticated model of the bulge/bar. Moreover, the data we use are only
sensitive to the bulge's contribution to radial forces. Therefore we do not
represent the bulge by a \df\ $f(\vJ)$ but by a fixed axisymmetric mass
distribution. We have updated our model from that used by BP15,
which followed \citet{McMillan2011} and was thus based on
\citet{Bissantz2002}. Our bulge model is now consistent with \citet{Wegg2013}
and \citet{Portail2015}. These authors  fitted dynamical models of the
bulge/bar  to near-IR photometry from the Vista
Variables of the Via Lactae survey \citep{Saito2012}
and line-of-sight velocity measurements of red-clump stars from
\cite{Ness2013}. 
They found a total mass within the cuboid with corners at
($\pm2.2,\pm1.4,\pm1.2\kpc$) to be $(1.84\pm0.07)\times10^{10}\mathrm{M}_\odot$.
Due to the axisymmetry of our model, we cannot reproduce this distribution
exactly but we construct our bulge to have the same mass within a similar
volume taking into account the mass of the discs and the dark halo. This
gives us a total mass of the bulge in this volume of
$1.05\times10^{10}\mathrm{M}_\odot$. We use a scale length for our
exponential that is the average of the two scale lengths in the $x$ and $y$
direction (the geometric mean is very similar).

\begin{table}
 \centering
 \caption{Parameters of the gas disc and the bulge. With the exception of $\Sigma_0$
 and $\Rd$, these parameters were fixed.}
 \label{tab:mass_model_params}
 \begin{tabular}{lcl}
  \hline\hline
  Parameter & value & unit \\
  \hline
  Gas disc\\
  $\Sigma_0$ & 	$0.38\Sigma_0(*)$&\\
  $\Rd$ &	$2\Rd({\rm stars})$&\\
  $z_\mathrm{d}$ &  $0.04$&$\kpc$\\
  $R_\mathrm{hole}$ & $4$&$\kpc$\\
  $M(\infty)$     & $17.7\times10^9$&$\msun$\\
  $M(R_0)$        & $3.53\times10^9$&$\msun$\\
  \hline
  Bulge\\
  $\rho_\mathrm{0,b}$	& $11.65$&$\msun\pc^{-3}$\\
  $r_\mathrm{0,b}$ & $0.56$&$\kpc$\\
  $r_\mathrm{cut,b}$& $2.1$&$\kpc$\\
  $q_\mathrm{b}$	&	 0.33& \\
  $\gamma_\mathrm{b}$	&	 0&\\
  $\beta_\mathrm{b}$	&	 1.8&\\
  $M(\mathrm{R}\lesssim2\,\mathrm{kpc})$     & $1.05\times10^{10}$&$\msun$\\
  $M(\inf)$     & $1.32\times10^{10}$&$\msun$\\
  \hline
 \end{tabular}
\end{table}

The density distributions of  the bulge is
\begin{equation} \label{eq:rho_spheres}
 \rho(R,z) = \frac{\rho_0}{m^\gamma(1+m)^{\beta-\gamma}}
 \exp[-(mr_0/r_\mathrm{cut})^2],
\end{equation}
where
\begin{equation}
 m(R,z) = \sqrt{(R/r_0)^2 + (z/qr_0)^2}.
\end{equation}
 Our model bulge has an axis ratio $q=0.33$, reflecting the ratio of
the scale length in the plane to that in the $z$ direction
\citep{Wegg2013}, and fades rapidly beyond $r_{\rm cut}=2.1\kpc$:
Table~\ref{tab:mass_model_params} lists all the parameters.

The gas disc is likewise represented by an axisymmetric distribution of
matter that has density
\begin{equation} \label{eq:rho_disc}
 \rho(R,z) = \frac{\Sigma_0}{2z_\mathrm{d}}
    \exp\left[-\left(\frac{R}{R_\mathrm{d}} + \frac{|z|}{z_\mathrm{d}} +
    \frac{R_\mathrm{hole}}{R}\right)\right].
\end{equation}
A non-zero value of the parameter $R_\mathrm{hole}$ creates a central cavity
in the disc.  The values of the parameters are given in
Table~\ref{tab:mass_model_params}.  The surface density normalisation is
adjusted to maintain the ratio $13.5:35.5$ between the gas and stellar
surface densities at $R_0$ that is given in \cite{Flynn2006}.  $\Rd$ and
$\Sigma_0$ are the only parameters that are varied: the other parameters are
fixed at the values adopted by P14 and earlier investigators.  As in previous
papers of this series and several other studies
\citep[e.g.][]{BovyRix2013,WeggGP2016} we set $\Rd({\rm gas})=2\Rd({\rm stars})$.
This setting may over-estimate $\Rd({\rm gas})$. However, the local surface
density of the gas disc is narrowly confined by the $13.5:35.5$ ratio given
above, and we have set $R_{\rm hole}=4\kpc$, so varying $\Rd({\rm gas})$ will
not have much impact on the structure of our model at $R<R_0$, which is what
is of concern here.

\subsection{Fitting algorithm}  
\label{sec:fitting}

\begin{figure}
\begin{center}
\includegraphics[width=\hsize]{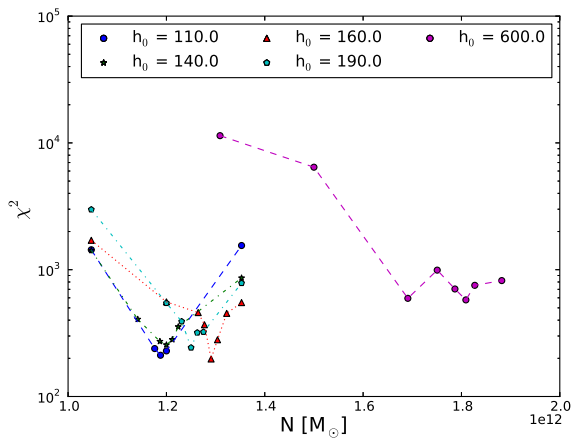}
\end{center}
\caption{Values of $\chi^2$ determined when searching for the optimum halo
normalisation $N$ given a value of the core action $h_0$.}\label{fig:logL}
\end{figure}

The algorithm we use to fit the DF to the data is essentially that used by
BP15 -- Table~\ref{tab:algorithm} lists its steps. 

Since the main differences
between our model and that of BP15 are in the inner Galaxy ($r\leq10\kpc$),
we fixed $J_0$ to their value, $J_0=6000\kpc\kms$. As they argue, this scale
action yields a realistic break radius $r_{\rm s}$ and has little bearing on the dark
halo's contribution to forces in the inner few kiloparsecs. This leaves $h_0$
and $N$ the only adjustable parameters of the dark halo's DF, one more than
the single free parameter, $N$, available to BP15. So for a grid of values of
$h_0$, we use the BP15 algorithm to determine $N$ and the nine free
parameters in the disc DF (two masses, four pseudo velocity dispersions, the
radial mass scale length and two radial scale lengths for the thick disc's
velocity dispersions). For every chosen value of $h_0$ we obtain a model that
satisfies all observational constraints other than the microlensing data. As
$h_0$ increases, the inner disc becomes steadily more massive so the optical
depth to microlensing increases. 

Fig.~\ref{fig:logL} shows for several values of $h_0$ the $\chi^2$ of the
\citetJuric\ data with respect to the model  vertical density profile
computed in step 10 of the fitting algorithm during the search for the
optimum value of the halo normalisation $N$. It is evident that for a given
value of $h_0$  remarkably
few values of $N$ have been tried, with the consequence that our accepted
value of $N$ almost certainly differs from the best possible value. The
sparseness of the search over $N$ reflects the computational cost of optimising
the disc DF for a given value of $N$: nine parameters can be adjusted, and
after each adjustment the self-consistent potential must be determined. The
potential is found by integrating the DF over three velocity components at
each node of a grid in the $Rz$ plane. Because the potential is determined
iteratively, these integrations have to be executed four or five times for
each choice of disc DF. Hence $\sim xx$ core-hours are required to optimise
$N$ for a given value of $h_0$. 

Since for some values of $h_0$ the adopted value of $N$ will come closer to
the true optimum value than for other values of $h_0$, the properties of the
recovered models do not change entirely systematically with $h_0$.  However,
the jitter is small enough to reveal systematic trends that would dominate if we
could always locate the true optimum value of $N$.

{\renewcommand{\arraystretch}{1.5}
\begin{table*}
\caption{Algorithm used to determine the ten remaining parameters of the
complete DF
given a value of $h_0$. }\label{tab:algorithm}
\begin{tabular}{cll}
\hline
Step&\hfil Description\hfil&\hfil Choose\hfil\\
\hline\hline
1& Choose values of the scale action $h_0$ and the parameter $N$ that controls the mass of the dark
halo.&\tabcol{$h_0,N$}\\

2&{\tabpar{Choose a pair of plausible double-exponential density distributions to
represent the stellar discs, and choose a plausible gas disc.}}
&\tabcol{$\Sigma_0,\Rd$, $z_{\rm d,thn},z_{\rm d,thk}$, $F_{\rm thk}$}\\

3& {\tabpar{Find the self-consistent equilibrium of the chosen dark halo in the
presence of the bulge, the gas disc and the  adopted models of the
stellar discs.}}
&\\

 4& {\tabpar{Adjust $\Sigma_0$ and $\Rd$ in the formulae for the stellar and
 gas discs until the constraints on $\Vc(R)$ listed in
   Section~\ref{sec:observ} are satisfied. Each time the disc parameters are
   changed, return to step 3 until changes become negligible.}}&\tabcol{$\Sigma_0,\Rd$}\\

5& {\tabpar{Choose a \df\ for the stellar disc
  that has the scale lengths found in Step 4 and adopt plausible
  values for its velocity-dispersion
  parameters.}}&\tabcol{$\Rd,\Sigma$, $R_{\sigma_R,{\rm
  thk}},R_{\sigma_z,{\rm thk}}$, $\sigma_{\rm
  R,thn},\sigma_{\rm z,thn}$, $\sigma_{\rm
  R,thk},\sigma_{\rm z,thk}$, $F_{\rm thk}$}\\

6& {\tabpar{At the nodes of a spatial grid, integrate the disc \df\ over
velocities to obtain the contribution to the density from the stellar disc.
After adding in the contributions from the gas, bulge and dark halo, solve for the associated
potential and adopt this potential.}}&\\

7&{\tabpar{At the nodes of a spatial grid, integrate the halo \df\ over
velocities to obtain the contribution to the density from the dark halo.
After adding in the contributions from the gas, bulge and stellar disc, solve
for the associated potential and adopt this potential.}}&\\

8&{\tabpar{If the update to the potential at Step 7 is non-negligible,
return to Step 6. Otherwise proceed to Step 9.}}&\\

9& {\tabpar{Adjust the velocity-dispersion parameters in the disc
\df\ to obtain a good fit of the model's kinematics to
the kinematics of the RAVE giants (including the contribution of the
stellar halo). Then return to Step 6 and continue  until changes to the
velocity-dispersion parameters become negligible.}}
&\tabcol{$\sigma_{\rm
  R,thn},\sigma_{\rm z,thn}$, $\sigma_{\rm
  R,thk},\sigma_{\rm z,thk}$, $F_{\rm thk}$}\\

10& {\tabpar{Compute the residuals between the vertical stellar
  density profile of \citetJuric\ and that implied by the
  {\df}s of the disc and stellar halo.}}&\\

11& {\tabpar{If these residuals are unsatisfactory, choose a new
normalisation for the
dark halo and return to Step 2. Otherwise, finish.}}&$N$\\
\hline
\end{tabular}
\end{table*} 
}

\section{Results}  
\label{sec:results}

\begin{table*}
  \caption[]{Properties of dark haloes generated by $\fDM$ defined in
Section~\ref{sec:DMDF}. All models have $J_0=6000\kpc\kms$ and
  $h_{\rm max}=10^6\kpc\kms$. $R_{\rho_{1/2}}$ 
    is the radius at which the DM density falls to half its central
    value. }
\label{tab:params}
  \begin{center}
    \begin{tabular}{lccccccc} 
      \hline
      &\multicolumn{7}{c}{$h_0/\kpc\kms$}\\
      \multicolumn{1}{l}{Parameter}&\multicolumn{1}{c}{$140$} &
      \multicolumn{1}{c}{$150$} &
      \multicolumn{1}{c}{$160$} &
      \multicolumn{1}{c}{$170$} &
      \multicolumn{1}{c}{$180$} &
      \multicolumn{1}{c}{$190$} &
      \multicolumn{1}{c}{$600$} \\

      \hline
\hline
 $\beta$ & 0.37 & 0.38 & 0.39 & 0.398 & 0.41 & 0.415 & 0.66 \\
 $N/10^{11}\msun$ & 12.0 & 12.6 & 12.9 & 12.0 & 13.5 & 12.5 & 18.1 \\
 $M_{200}^{\rm DM}/10^{11}\msun$ & 7.51 & 7.97 & 8.20 & 7.54 & 8.67 & 7.95 & 12.7\\
 $R_{200}/\kpc$ & 183 & 186 & 188 & 183 & 191 & 186 & 217 \\
$R_{\rho_{1/2}}/\kpc$ & 1.90 & 1.97 & 2.03 & 2.11 & 2.15 & 2.18 & 3.59 \\
$\rho($R$_0,0)/\msun\pc^{-3}$ & 0.0126 & 0.0131 & 0.0133 & 0.0120 & 0.0139 & 0.0124 & 0.0109 \\
$\rho($R$_0,0)/\GeVc\cm^{-3}$ & 0.477 & 0.497 & 0.504 & 0.456 & 0.527 & 0.472 & 0.415 \\
\hline
    \end{tabular}
  \end{center}
\end{table*}

\begin{table*}
  \caption[]{Parameters of the DFs of the stellar discs. The model listed
  under $h_0=0$ is that recovered by BP15. The quoted surface densities are
  (i) the surface density of stars, and (ii) the surface density of all
  matter, both within $1.1\kpc$ of the plane.}
\label{tab:discs}
  \begin{center}
    \begin{tabular}{lccccccccc} 
      \hline
      &\multicolumn{8}{c}{$h_0/\kpc\kms$}\\
      \multicolumn{1}{l}{Parameter}&
      \multicolumn{1}{c}{$0$ (NFW)} &
      \multicolumn{1}{c}{$140$} &
      \multicolumn{1}{c}{$150$} &
      \multicolumn{1}{c}{$160$} &
      \multicolumn{1}{c}{$170$} &
      \multicolumn{1}{c}{$180$} &
      \multicolumn{1}{c}{$190$} &
      \multicolumn{1}{c}{$600$} \\
      \hline
\hline
 Thin and thick disc   &  &  &  &  &  &  \\
 $\Rd/\kpc$ & 3.66 & 3.08 & 2.97 & 2.92 & 2.95 & 2.80 & 3.00 & 2.73 \\
 $M_\rd/10^{10}\msun$ & 3.6 & 3.91 & 3.81 & 3.76 & 4.26 & 3.63 & 4.29 & 5.39\\ 
$\Sigma_0(*)/\msun\pc^{-2}$&43.7 &45.6&43.0&41.7&50.0&38.6&49.1&55.5\\
$\Sigma_0({\rm tot})/\msun\pc^{-2}$&88.3  &89.9&87.3&85.8&94.3&82.8&94.2&99.4\\
\hline
Thin disc  &  &  &  &  \\
$\sigma_\mathrm{R,thn}/\kms$  & 35.40 & 33.59 & 33.42 & 33.87 & 33.38 & 33.00 & 34.24 & 31.61 \\
$\sigma_\mathrm{z,thn}/\kms$  & 26.00 & 25.70 & 25.67 & 26.61 & 26.40 &
25.54 & 27.05 & 25.43 \\
$\mathrm{R}_\mathrm{\sigma_R,thn}/\kpc$  & 2$R_\rd$ & 2$R_\rd$ & 2$R_\rd$ & 2$R_\rd$ & 2$R_\rd$ & 2$R_\rd$ & 2$R_\rd$ & 2$R_\rd$ \\
$\mathrm{R}_\mathrm{\sigma_z,thn}/\kpc$  & 2$R_\rd$ & 2$R_\rd$ & 2$R_\rd$ & 2$R_\rd$ & 2$R_\rd$ & 2$R_\rd$ & 2$R_\rd$ & 2$R_\rd$ \\
\hline
Thick disc  &  &  &  &  &  &  &  \\
$\sigma_\mathrm{R,thk}/\kms$  & 52.78 & 51.31 & 51.68 & 51.93 & 51.18 & 51.97 & 47.59 & 49.75 \\
$\sigma_\mathrm{z,thk}/\kms$  & 53.33 & 51.89 & 51.65 & 53.97 & 53.82 & 51.27 & 53.96 & 50.25 \\
$\mathrm{R}_\mathrm{\sigma_R,thk}/\kpc$  & 11.6 & 12.90 & 13.41 & 12.47 & 12.73 & 14.59 & 11.50 & 12.70 \\
$\mathrm{R}_\mathrm{\sigma_z,thk}/\kpc$ & 5.01 & 4.47  & 4.29 & 4.40 & 3.92  & 4.48 & 4.67 & 4.35 \\
$\mathrm{F}_\mathrm{thk}$  & 0.416 & 0.465 & 0.444 & 0.402 & 0.478 & 0.455 & 0.452 & 0.488 \\
\hline
    \end{tabular}
  \end{center}
\end{table*}

Tables~\ref{tab:params} and \ref{tab:discs} present, respectively, the
parameters of the dark haloes and the stellar discs of six models that are
consistent with all the data, including the microlensing optical depths --
these are the models with $h_0$ in the range $(140,190)\kpc\kms$. The last
column in each table describes a model, described in
Section~\ref{sec:nogood}, that has $h_0=600\kpc\kms$ and is not satisfactory.
Table~\ref{tab:params} shows that in the six satisfactory models, the radius
$R_{\rho1/2}$ at which the density of the dark halo falls to half its central
value increases with $h_0$ from $1.9$ to $2.2\kpc$. The local DM
density varies in the range
$(0.0120,0.0139)\,\mathrm{M}_\odot\,\mathrm{pc}^{-3}$
[$(0.456,0.527)\,\mathrm{GeV} \mathrm{cm}^{-3}$]. The value of $\rho_{\rm DM}(R_0,0)$
does not vary systematically with $h_0$ and we believe the width of the given range
reflects the uncertainty in the data and has no physical significance.
Table~\ref{tab:discs} shows that the scale lengths of the stellar discs vary
in the range $(2.80,3.08)\kpc$, again without systematic dependence on $h_0$.
Hence all satisfactory models have disc scale lengths that are significantly
shorter than the value $R_\rd=3.66\kpc$ recovered by BP15 --
Table~\ref{tab:discs} gives all the parameters of that disc in the column for
$h_0=0$.  When $h_0$ is significantly non-zero, there is less DM at
$r\la2\kpc$, so, given that the surface density at the Sun is essentially
fixed by the \citetJuric\ and RAVE data, the disc needs to have a shorter
scalelength to compensate. By the same token, the mass of the stellar disc
must be larger than when $h_0=0$. Indeed, in the new models, the total
stellar disc mass lies in the range $(3.63,4.29)\times10^{10}\msun$ with the
mass inside the solar circle in the range $(3.02,3.53)\times10^{10}\msun$.
When we include the bulge and gas disc, the total baryonic mass lies in the
range $(6.7,7.4)\times10^{10}\msun$. Hence, at 7.3 to 8.8 per cent our
baryonic fractions are significantly higher than that, 4.2 per cent, of BP15.

\begin{figure}
\begin{tabular}{ll}
\includegraphics[width=1.05\hsize,angle=0]{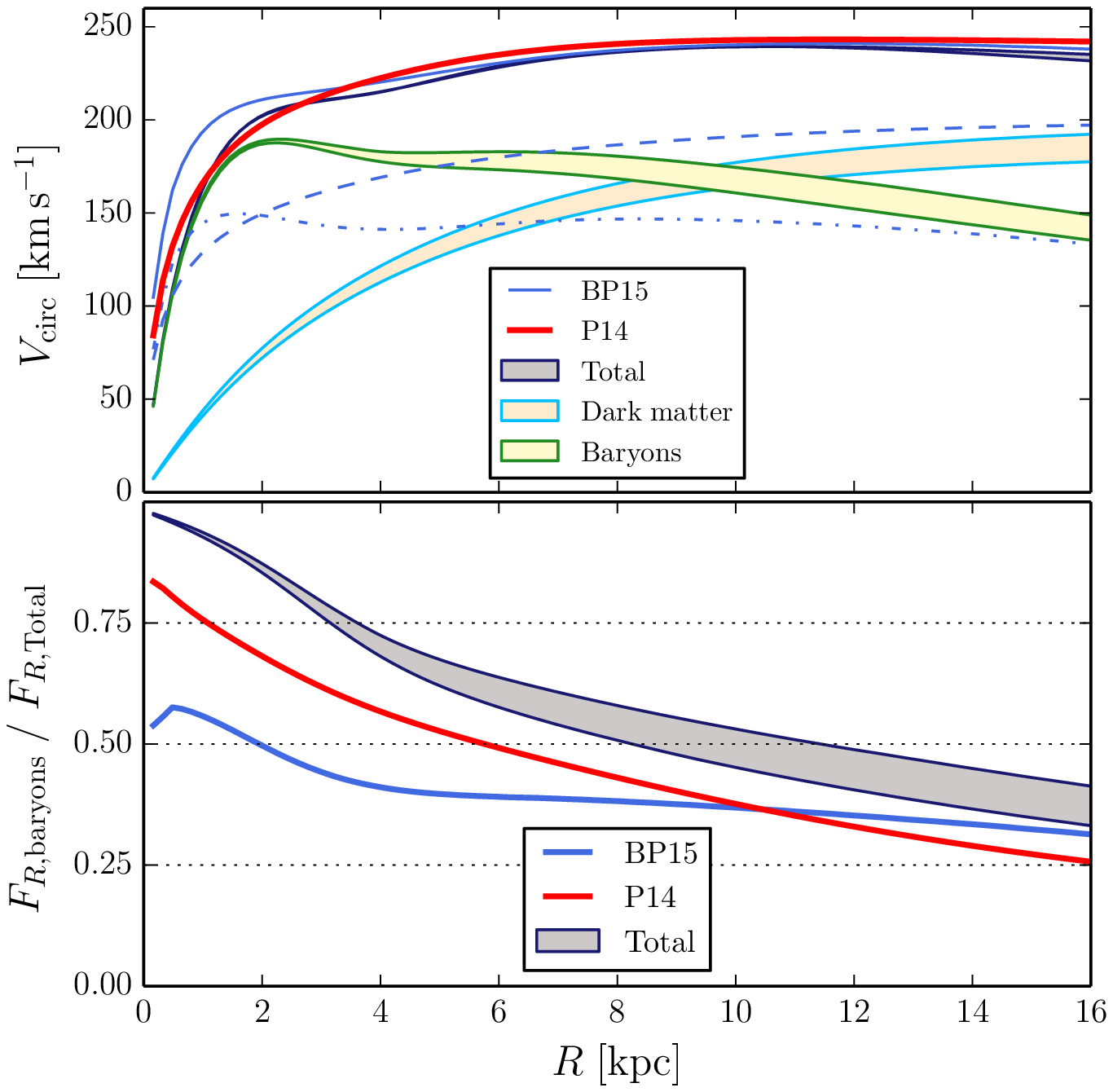}
\end{tabular}
\caption{Upper panel: $\Vc(R)$ for one of our Galaxy models (black
  line) compared to the models of BP15 (red lines) and
  P14 (blue lines). The range of dark halo and baryonic
  contributions are shown as filled areas outlined in blue (dark halo)
  and green (baryonic). The blue dashed line is the dark halo
  contribution to the rotation curve of BP15 and the
  green dash-dotted line is its baryonic component.  Lower panel: the
  ratio of radial forces from the baryonic component to the total mass
  distribution for the same models shown above. We use no observational
  constraints on $\Vc$ at $R<R_0/2$.}
\label{fig:rotfrat}
\end{figure}

In the upper panel of Fig.~\ref{fig:rotfrat} we plot the circular speed
curves of several models: the circular-speed curves of the new models lie
within the very narrow shaded area that is bounded by black lines. The yellow
and orange shaded areas show the circular speeds that are generated by the
baryons and DM, respectively, in the new models. Naturally, these
individual curves cover wider range than the total, since $\Vc$ from the
baryons rises with $h_0$ and this rise is largely compensated by a fall in
$\Vc$ from the dark halo. Within the uncertainties, the contributions to
$\Vc$ become equal at $R_0$.

The blue curve shows $\Vc$ in the model of P14 that has axis ratio $q=0.8$;
it differs insignificantly from the dark shaded region of the new models. The
red curve shows $\Vc$ from BP15.  It rises more steeply at small radii than
the other curves but is consistent with them in the region $R>4\kpc$ for
which we have constraints.  The broken curves show the circular speeds
generated by the dark halo (blue) and the baryons (dark green) in the BP15
model. They differ strongly from the corresponding curves of the new models,
in particular crossing at $R<2\kpc$ rather than at $R>8\kpc$.

In the lower panel of Fig.~\ref{fig:rotfrat} we plot the fraction of the
total radial force that is contributed by the baryons. In the new models
(dark shaded region) this is naturally larger than in the models of either
P14 or BP15. In particular, at $R_0$ the baryons contribute $0.53\pm0.03$ of
the radial force in the new models.

\begin{figure*}
\begin{center}
\includegraphics[width=\hsize,angle=0]{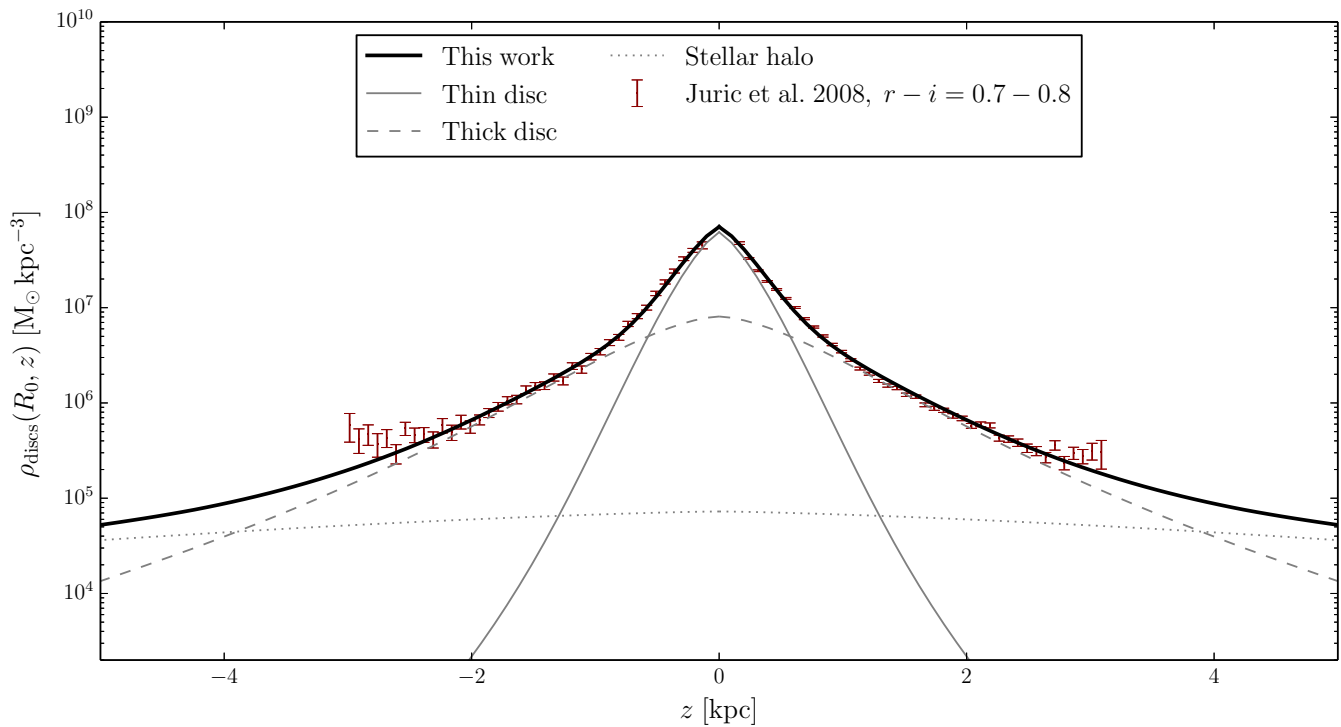}
\end{center}
\caption{Vertical stellar density profiles in the solar annulus of our
   model with $h_{0}=150\kpc\kms$ showing total density (solid line), the
  thin disc(dashed line), the thick disc(dashed line) and the stellar
  halo (dotted line). Star count data from \citetJuric\ used in
  the fitting are shown as red error bars.}
\label{fig:vdensprf}
\end{figure*}

The curves in Fig.~\ref{fig:vdensprf} show the vertical density profiles of
the thin and thick discs and the stellar halo above the Sun in the model with
$h_{0}=150\kpc\kms$. The data points show the \citetJuric\ data, and we see
that they agree well with the summed density profile for this model. The other models fit the \citetJuric\
data with comparable precision.

\begin{figure*}
\begin{center}
\includegraphics[width=\hsize,angle=0]{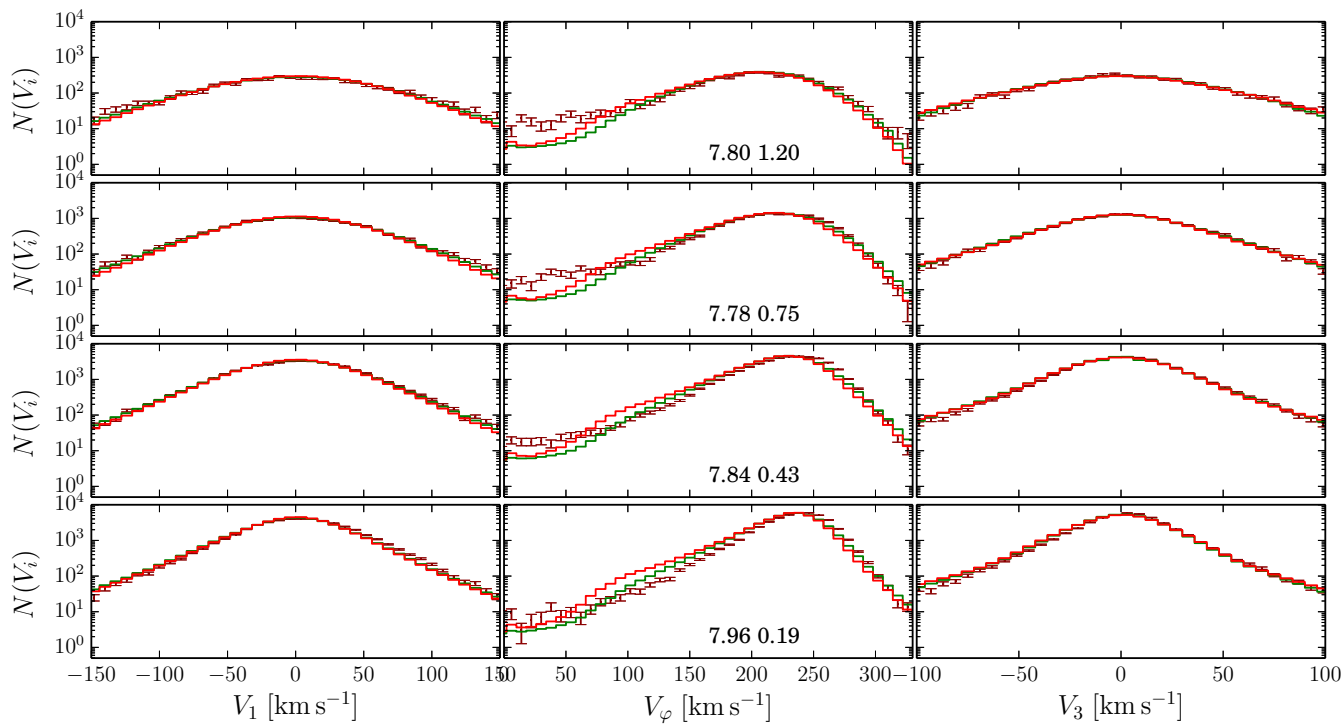}
\end{center}
\caption{Histograms of three orthogonal components of velocity for four
spatial bins bounded by $R=R_0$ and $R=R_0-1\kpc$. The $(R,z)$ coordinates of
the barycentre
of each bin are given at the bottom of each central panel. The red curves
show the predictions (after simulating errors in proper motion and distance)
of the model with $h_0=150\kpc\kms$. The green curves show the same
quantities for the BP15 model, while the data points show the observed
histograms of the giants in RAVE. The velocity components are the principal
axes of the velocity ellipsoids computed  by Binney et al.\ (2014), but to a
good approximation $V_1=v_R$ and $V_3=v_z$.}
\label{fig:RAVE}
\end{figure*}

The red lines in Fig.~\ref{fig:RAVE} show the fit to the kinematics of the
RAVE giants (data points) achieved by the model with $h_0=150\kpc\kms$ in the
four spatial bins that lie inside $R_0$. The left and right columns show that
the model provides fits of
outstanding quality to the distributions or radial and vertical velocities.
The fit to the distributions of $v_\phi$ components is good, if not quite as
good as that obtained by BP15 (green lines). The most significant
discrepancies are under-provision of stars with large $v_\phi$ further than
$1\kpc$ from the plane (top central panel). This deficit suggests that the
scale length of the thick disc should be longer. In fact, \citetJuric\
reported $R_\rd=3.6\kpc$ for the thick disc and $R_\rd=2.6\kpc$ for
the thin disc, whereas we have assumed from the outset that the two discs
have the same scalelengths. 

However, evidence is accumulating that the disc formed by $\alpha$-high stars
has a smaller value of $R_\rd$ than the disc formed by the $\alpha$-low stars
\citep{Hayden2015}, so if one identifies the former disc with the thick disc, one
has a conflict with \citetJuric. The likely resolution of this conflict is
that the disc formed by the $\alpha$-low stars flares, so the scalelength of
{\it all} stars that lie at $|z|>0.7\kpc$ is in fact larger than the
scalelength of the stars that lie at $|z|<0.3\kpc$. This would explain the
deficit in our models of stars with large $v_\phi$ revealed by the central
top panel of Fig.~\ref{fig:RAVE}.

A significant, but less serious, problem in Fig.~\ref{fig:RAVE} is
under-provision of stars with low $v_\phi$ in the central top panel. This
problem arises also in the model of BP15 and signals weakness in our choice
of the DF of the stellar halo.

Our model provides a better overall fit to the data further from the plane
(upper two centre panels) but has an excess of non-rotating stars at low
values of $z$.

\begin{figure}
\begin{tabular}{ll}
\includegraphics[width=1.05\hsize,angle=0]{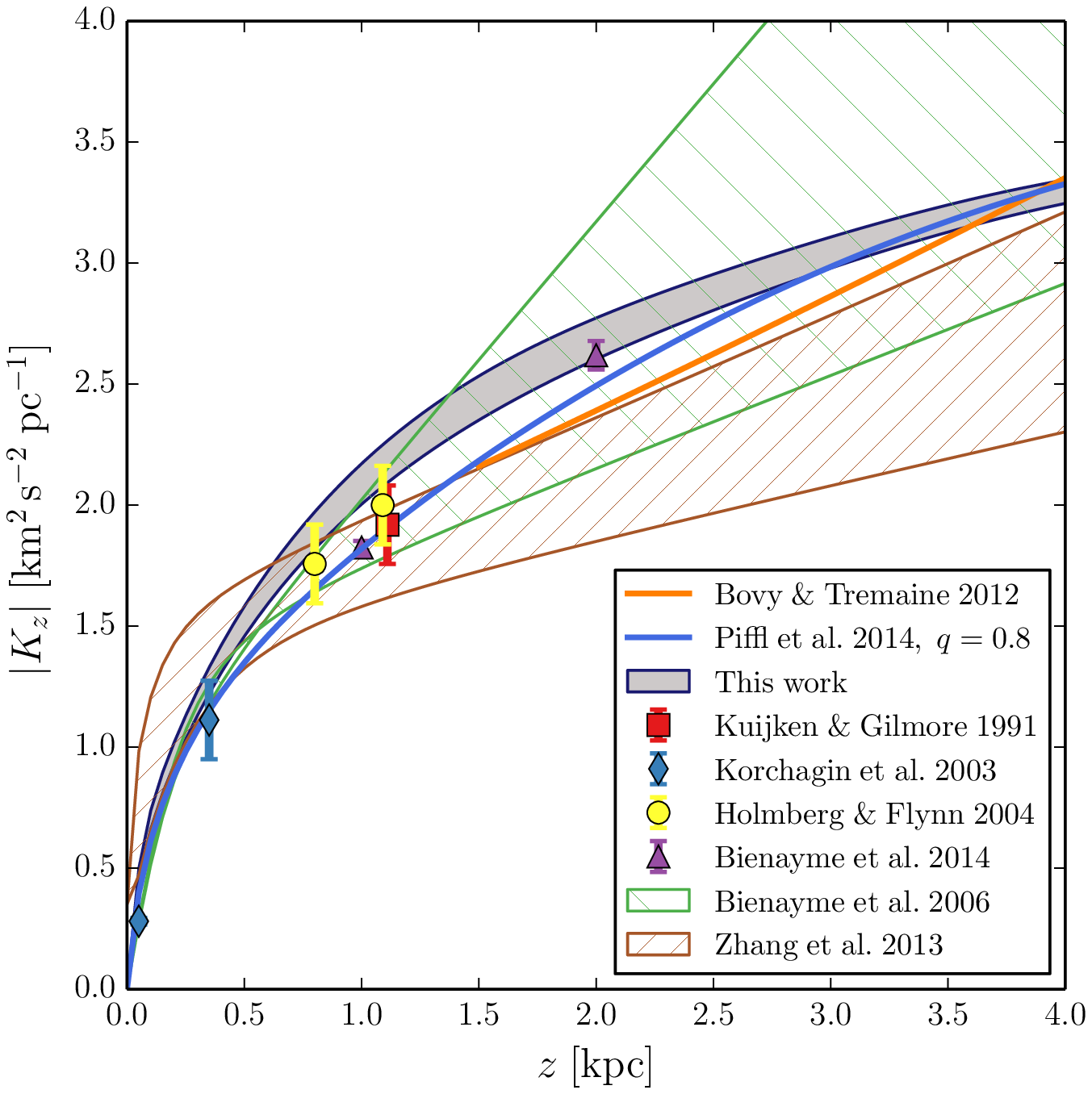}
\end{tabular}
\caption{Vertical force $|K_z|$ in the solar cylinder as a function
  of distance $z$ from the Galactic plane. The new models have $K_z$ in the
  grey shaded region.}
\label{fig:Kz}
\end{figure}

In Fig.~\ref{fig:Kz} the dark shaded region shows the range of values of the
vertical force $|K_z|$ provided by the new models. This lies slightly above
the $K_z$ curve from P14 (blue curve) because the new discs have slightly
higher stellar surface densities at $R_0$ ($41.7-50.0\msun\pc^{-2}$ Table~\ref{tab:discs}) than the
P14 disc ($37.1\msun\pc^{-2}$). Nonetheless, the new
models are consistent with the classic determination of \cite{KuijkenG}.
They are also consistent with the findings of \cite{BienaymeRAVE}, indicated
by green boundary curves.

\subsection{Structure of the dark halo}  
\label{sec:DMmass}

\begin{figure}
\includegraphics[width=1.1\hsize,angle=0]{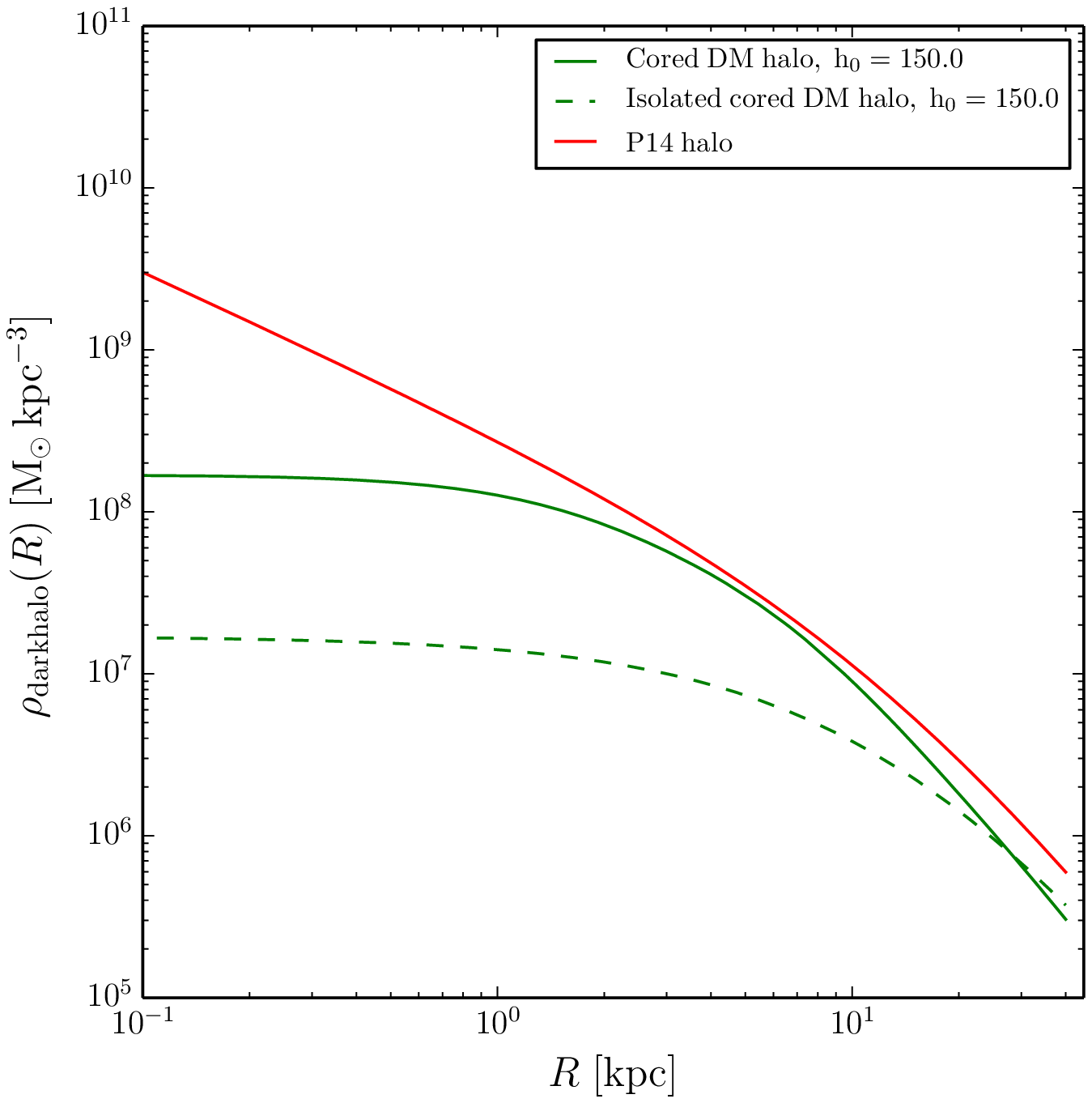}
\caption{The red curve shows the  density profile within the equatorial
plane of the dark halo of the $q=0.8$ halo of P14.
The solid green curve shows the  corresponding
density profile of
the model with $h_0=150\kpc\kms$, while the broken green curve shows the
density profile of that halo when the baryons are removed.}\label{fig:twohalos}
\end{figure}

The green curves in Fig.~\ref{fig:twohalos} show the density profiles in the
equatorial plane of the model with $h_0=150\kpc\kms$ (i) in the
full model (full curve), and (ii) after removal of the baryons (dashed curve).
We see that the gravitational field generated by the baryons increases the
halo
density by an order of magnitude in the region $r\lesssim3\kpc$, but changes
it very little at $r\gtrsim20\kpc$. In the full model the density of the dark
halo at $r=R_0$ is almost the same as in the P14 model (red curve). The density of the
dark halo falls well below that of the P14 model only at
$r\lesssim1\kpc$.

At $r\gtrsim R_0$ the density profile of our halo rises more steeply than
that of an NFW halo because the inward pull of the baryons' gravitational
field is centrally concentrated. Thus at $r>20\kpc$ the two green curves
Fig.~\ref{fig:twohalos} asymptote to one another and lie below but parallel
to the red curve of an NFW halo. On account of the inward pull of the
baryons, the full green curve almost touches the red curve around $R_0$
because the density of DM at $R_0$ is tightly constrained by the
data, and the density everywhere else follows from this density and the
structure of our chosen DF.

The third row of Table~\ref{tab:params} gives the values of the mass,
$M_{200}$, of DM inside the radius where the halo's mean density is
200 times the mean cosmic density of matter. At
$M_{200}\sim0.8\times10^{12}\msun$ these values are significantly less than
the value, $M_{200}=(1.3\pm0.1)\times10^{12}\msun$ for the P14 model and that,
$M_{200}\sim1.4\times10^{12}\msun$, found by BP15. This reduction in
$M_{200}$ is a straightfoward consequence of the dynamics described in the
previous paragraph.  However, there are precedents for such
light dark  haloes: \citet{Penarrubia2014}
estimate the Milky Way's dark halo mass as
$0.8^{+0.4}_{-0.3}\times10^{12}\msun$ and a recent estimate by
\cite{Huang2016} gives a halo mass of
$0.9^{+0.07}_{-0.08}\times10^{12}\msun$.

Our haloes do yield local DM densities that are higher than many
encountered in the literature \citep[see the review of][]{Read2014}. In
particular, the dark haloes of \cite{WeggGP2016} yield local densities in the
range $(0.005-0.008)\msun\pc^{-3}$. However, from the fact that
\cite{BienaymeRAVE} derived a very similar value to us from a completely
different analysis of the RAVE data, we conclude that our large local
DM density follows from the unprecedented RAVE data and supersedes
earlier estimates.

\section{Optical depths to microlensing}  
\label{sec:mulens}

In this section we investigate which models are consistent with the
microlensing data. The latter were for many years subject to a regrettable
level of uncertainty on account of the phenomenon of `blending': stars that
will be lensed merging with brighter stars with the result that one
under-estimates the number of stars being monitored and thus over-estimates
the optical depth to lensing \citep{WozniakPaczynski1997}. Recognition of this phenomenon
shifted attention to the statistics for lensing of apparently bright stars,
such as red clump stars, which are less affected by blending
\citep{Popowski2001,Popowski2005}. However, it now
seems that blending is well enough understood for valid optical depths to be
determined from the lensing statistics of all stars, not just the brightest
ones \citep{Sumi2016}. In particular, Sumi et al., analysing data from the
MOA-II survey,\footnote{www.massey.ac.nz/$\sim$iabond/alert2000/alert.html} find similar values of the optical depth from all stars and
from red-clump stars. Below we compare the optical depths of our models to
the data in \cite{Sumi2016}.

\cite{WeggGP2016} have recently evaluated the optical depth to microlensing
predicted by their Galaxy models. Inside $R_{\rm cut}\simeq1.6\kpc$ these
models comprise an N-body model from \cite{Portail2015}, while at $R>R_{\rm
cut}$ the models comprise a double-exponential stellar disc -- the latter
consists of a thin disc only with local surface density
$\Sigma_*(R_0)=38\msun\pc^{-2}$.  In some cases this disc has scale height
$\zd=0.3\kpc$, and in others it flares from $\zd=0.18\kpc$ at $R=4.5\kpc$ to
$\zd=0.3\kpc$ at $R_0$. Given that our bulge is an axi-symmetrised version of
a bulge/bar from \cite{Portail2015}, these models are very similar to ours. 

When viewed close to its major axis, a barred Galaxy will have a larger optical depth
to microlensing, $\tau$, than the axisymmetric model with the same circular-speed
curve. As \cite{WeggGP2016} point out, the angle, $\sim25^\circ$, between the
long axis of the bar and the Sun-Centre line is such that we expect an
axisymmetric model to under-estimate $\tau$ only slightly. Most
lenses will lie in the region $R>R_{\rm cut}$ where \cite{WeggGP2016}
assume axisymmetry, so their optical depths will be under-estimated to a
similar extent to ours.

\cite{WeggGP2016}  compute $\tau$ for each of the MOA-II fields taking into
account the luminosity function of source stars. Together with the apparent
magnitude limit of the survey, the latter determines the distribution along
the line of sight of the surveyed stars, and therefore the mean optical
depth in the given field. Here we perform a much simpler calculation: we
compute
\begin{equation} \label{eq:mod}
\tau=\frac{4\pi{G}}{c^2}\int_{0}^{S}\rd D\,\rho(D)\frac{D(S-D)}{S},
\end{equation}
for a star that is located at $(R,z)=(0,R_0\tan b)$. Here $S=R_0\sec b$ is
the distance to the source, $D$ is the distance to the deflector and
$\rho(D)$ is the mass density contributed by the deflectors. The value of
$\tau$ from equation (\ref{eq:mod}) is typical of what the more sophisticated
computation yields after averaging over sources fainter than $I_0\simeq12$
\citep[see][Fig.~1]{WeggGP2016}. 

\begin{figure}
\begin{center}
\includegraphics[width=1.1\hsize,angle=0]{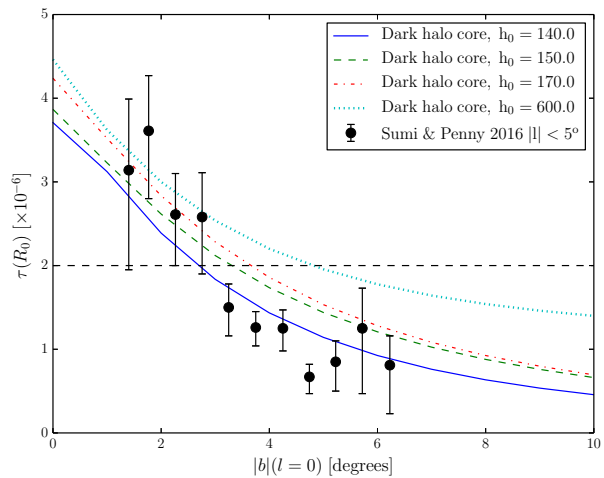}
\end{center}
\caption{The optical depth to microlensing for a star located in the new models  at
$(R,z)=(0,R_0\tan b)$. The data points are from Sumi \& Penny (2026) for
fields that have $|l|<10^\circ$. Each curve gives the prediction of a model
at $l=0$.}
\label{fig:microl_h0}
\end{figure}

The curves in Fig.~\ref{fig:microl_h0} show the predicted optical depths.
As expected, at any given $b$ the optical depth increases with $h_0$. As
$h_0$ increases the decline in $\tau$ with $b$ becomes less steep, and the
steepest curves are most consistent with the data points. The curve for
$h_0=600\kpc\kms$ is certainly too shallow.

BP15 argued that their model was inconsistent with the
measured optical depth to microlensing of bulge stars because in it the mass
of stars inside $R_0$ was less than the 
$\sim3.9\times10^{10}\msun$ considered essential by \cite{Binney2001}. Our
bulge model has $\sim1.05\times10^{10}\msun$ at $R<2\kpc$, so by the 
criterion in Binney \& Evans, the stellar disc should contain $\sim2.85\times10^{10}\msun$
within $R_0$. In the new  models the mass of the stellar disc within $R_0$
ranges from  3.02 to $3.53\times10^{10}\msun$ within $4\kpc$ of the
galactic plane and between 2.55 and $2.97\times10^{10}\msun$
within $1\kpc$ of the galactic plane, so they just about satisfy the
criterion of Binney and Evans.

\section{An over-large core}
\label{sec:nogood}

A minimum value of $h_0$ is set by the requirement for sufficient
microlensing optical depth. Now we show what happens when $h_0$ is made too
large. 

A large value of $h_0$ implies a large core radius of the dark halo. A
compact disc is then required to keep the circular speed quite flat given a
radially rising contribution from the dark halo. Since the disc's surface
density at $R_0$ is constrained by the RAVE data, a disc with a small value
of $\Rd$ is a massive disc. The final column of Table~\ref{tab:discs}
quantifies these points by showing that $h_0=600\kpc\kms$ yields
$\Rd=2.73\kpc$ and $\Md=5.39\times10^{10}\msun$. The final column of
Table~\ref{tab:params} shows that with with $h_0=600\kpc\kms$ the local
DM density $\rho_{\rm DM}(R_0)$ is only $0.010\msun\pc^{-3}$ yet the
reference mass $M_{200}=12.7\times10^{11}\msun$ is more than 50 per cent
higher than in the acceptable models. With a large core in the halo, a
sufficiently large value of $\rho_{\rm DM}(R_0)$ can only be achieved
alongside a  large density at $r\gg R_0$. For this reason,
$\rho_{\rm DM}(R_0)$ is depressed by large $h_0$.

\begin{figure}
\begin{center}
\includegraphics[width=1.1\hsize,angle=0]{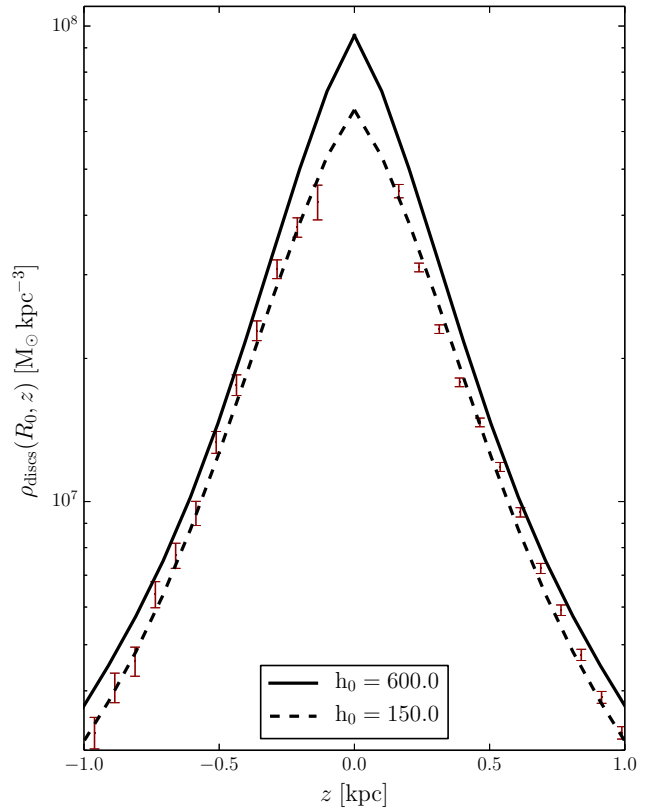}
\end{center}
\caption{Comparison of the vertical stellar profiles at $R_0$ of an
acceptable model ($h_0=150\kpc\kms$, broken curve) and a model with
over-large $h_0$ (full curve). The \citetJuric\ data points can be slid up
or down at will.}
\label{fig:too_large_rho}
\end{figure}

\begin{figure}
\begin{center}
\includegraphics[width=1.1\hsize,angle=0]{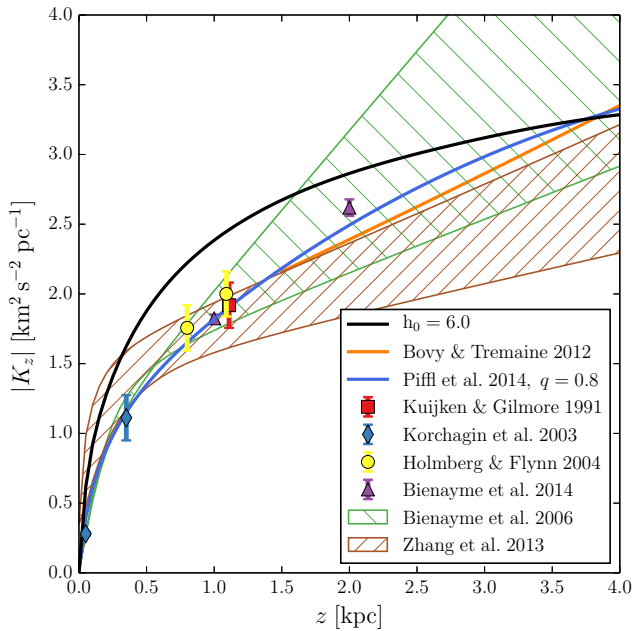}
\end{center}
\caption{The full curve shows $K_z$ for a model in which $h_0=600\kpc\kms$ is
unacceptably large.}
\label{fig:too_large_Kz}
\end{figure}

Fig.~\ref{fig:too_large_rho} compares the stellar density profiles above the
Sun in an acceptable model ($h_0=150\kpc\kms$) and the model with over-large
$h_0$. The full curve for the model with over-large $h_0$ lies above that of
the acceptable model. This is not a problem because the data points can be
shifted up or down by increasing or decreasing the mass-to-light ratio of the
stars. What {\it is} an issue is that the full curve is steeper at
$|z|\lesssim0.5\kpc$ because the model with over-large $h_0$ has the a more
massive
disc. As a consequence of this change of shape, by sliding the data points
vertically, they can be brought into closer agreement with the broken than
with the full curve. The black curve in Fig.~\ref{fig:too_large_Kz} shows
that the massive thin disc of the model with $h_0=600\kpc\kms$ generates
values of $K_z$ that are significantly larger than most estimates.

\section{Discussion}\label{sec:discuss}

In the CDM paradigm, structure formation commences with DM at an essentially
infinite phase-space density. As the DM aggregates gravitationally, the
fluctuating gravitational potential that the DM particles experience,
scatters particles to lower phase-space densities. Remarkably, simulations of
this process revealed that extremely high, possibly infinite phase-space
densities of DM survive at the centres of dark haloes. No analytic function
provides a perfect representation of the central structures of the haloes
that form in DM-only simulations, but the NFW profile and the Einasto profile
\citep{Einasto1965} both provide accurate fits \citep[e.g.][]{DuffySchaye}. Whereas the NFW
profile requires an infinite central density, as $|\vJ|\to0$ the phase-space
density associated with the Einasto profile rises steeply to finite value.
Even in the Einasto model, however, in no part of action space is the
phase-space density essentially constant. We consider this a weakness because
on general physical grounds, scattering will create such a region around any
peak in the phase-space density.

Numerical experiments have shown that dark haloes have cuspy cores regardless
of whether they form primarily through violent relaxation of single large
structures, or through repeated merging of many small haloes
\citep{HussJainSteinmetz1999,MooreQGSL}. This
finding implies that when a small halo is cannibalised by a larger halo,
particles released at high phase-space density by the small halo rather
precisely replace in action space the particles of the larger halo that have
been kicked up to higher actions.

\cite{ElZant} pointed out that when a baryonic structure falls into a
dark halo, the scattering of the host halo's particles to regions of
lower phase-space density is not compensated by release of DM
particles by the inspiralling structure, and the central cusp of the
host halo will be eliminated. \cite{JardelSellwood2009} quantified
this proposal using N-body simulations of spherical systems, and
reported that the effect is weaker than \cite{ElZant} predicted. They
did find, however, that when $0.1M_{200}$ is distributed amongst 500
clumps that initially orbit in the halo out to $4r_{\rm s}$, after
$9\Gyr$ a core develops that extends out to $\sim r_{\rm s}/50$, which
corresponds to $\sim0.5\kpc$ in the case of our Galaxy.  Reducing the
number of clumps from 500 causes the scale of the core to
grow. \cite{Cole2011} investigated clumps on a variety of initial
orbits and found that a clump with mass 1 per cent of the mass of the halo
could remove about twice its own mass from the inner halo,
transforming a cusp into a core or weak cusp. This effect was
enhanced if the clump was subsequently removed, as by a galactic
wind.

\cite{HernquistWeinberg1992} pointed out that a bar rotating within a
spheroidal component will experience dynamical friction as a consequence of
upscattering the spheroid's particles. Subsequent N-body simulations of bars
in spheroids \citep{DebattistaSellwood,Athanassoula2003} have shown
that this is an important effect: the bar responds to the loss of energy and
angular momentum to the spheroid by becoming longer, slower and stronger. The
response of the spheroid to the bar is less well documented.
\cite{WeinbergKatz2002} showed that a bar would tend to produce a core in a
dark halo.  \cite{HolleyB2005} confirmed that a bar will produce at least a
small core in its dark halo, but \cite{Sellwood2003} showed that at small
radii the DM density nevertheless rises as the disc forms because at
most radii compression by the growing gravitational field of the baryons
overwhelms the effect of core formation. This finding underlines  the
important role played by galactic winds in expelling baryons from the central
galaxy after then have upscattered DM.   

We believe our proposal is consistent with all these studies. In particular,
our favoured core radii are much smaller than $r_{\rm s}$ and, as
Fig.~\ref{fig:twohalos} illustrates, upscattering has probably not fully
compensated for adiabatic compression through most of the star-forming disc.

\cite{BinneySG2001} argued that  massive outflows from early galaxies
indicate that the centres of galaxies have processed significantly more
baryons than now reside in visible galaxies. As baryons sank towards the
centres of a dark halo, they probably surrendered a considerable amount
of angular momentum to the dark halo, causing it to expand. Later  stellar
feedback pushed the baryons out into the circum/intergalactic medium, where
they now reside. The case for ejection of large quantities of baryons is now
established but the impact of these baryons on the dark halo remains
uncertain. In a similar spirit 
\cite{NipotiBinney2015} argued that early in the life of a halo, baryons are
likely to accumulate in the form of gas until their gravitational field has
comparable strength to that of the DM. At that point the body of gas
in a low-mass dark halo
is likely to break up into a small number of blobs, each of which contains a
significant fraction of the total mass interior to its orbit. In these
circumstances, a blob will dump its orbital energy into
DM within a couple of dynamical times. The take-up of this energy
will significantly reduce the central phase-space density of DM.

\cite{NavarroEkeFrenk} investigated the effect of baryon ejection on a dark
halo in the axisymmetric limit and found it to be a significant process, but
\cite{GnedinZhao2002} concluded from a study of the spherical limit that the
effect is too weak to be of interest. In our view angular-momentum transfer,
which is suppressed in the axisymmetric case and eliminated in the spherical
case, lies at the core of the process, so these studies are of marginal value.

We conclude that some combination of a bar and massive baryonic lumps are
likely capable of smoothing the cusp in the phase-space density at the centre
of a dark halo out to $\sim r_{\rm s}/15$ as our analysis suggest has
happened in our Galaxy.

\section{Conclusions}\label{sec:conclude}

BP15 showed that an NFW DF is inconsistent with data for our Galaxy,
presumably because the most tightly bound DM particles have ben scattered to
a nearly constant phase-space density by fluctuations in the overall
gravitational potential. Hence we have here introduced a three-parameter
family of analytic functions $f(\vJ)$ as candidates for the DFs of dark
haloes in galaxies like ours. One parameter controls the halo's scale radius
$r_{\rm s}$, another controls its mass, and the third, $h_0$, controls the
size of the region around the origin of action space in which the phase-space
density of DM is almost constant. The existence of this region distinguishes
these haloes from standard NFW haloes, in which the phase-space density
diverges as $|\vJ|\to0$, and from  Einasto haloes, in which the
phase-space density rises to a sharp peak as $|\vJ|\to0$. 

We have searched the family of new halo DFs and our usual family of stellar DFs for
sets of parameters that self-consistently generate a Galaxy model that is
consistent with constraints on the rotation curve, the vertical structure of
the stellar disc in the solar cylinder and the kinematics of giants in RAVE.
There is a range of values of $h_0$ for which these constraints are satisfied
and optical depths to microlensing of bulge stars are predicted that are
consistent with the results of surveys for microlensing events. In the new
models the local DM density [$\rho_{\rm
DM}(R_0,0)=(0.0131\pm0.0007)\msun\pc^{-3}$] and stellar surface density
[$\Sigma_{\rm B}(R_0)=45\pm 4\msun\pc^{-2}$] are very similar to the values
determined by P14 and BP15, but with $M_\rd=(3.95\pm0.3)\times10^{10}\msun$
the new discs are more massive than those of BP15 because the new models have
shorter disc scale lengths ($R_\rd\la3\kpc$) than the BP15 model. The haloes
have core radii $2\pm0.05\kpc$. The microlensing data are critical for these
fits because they place a lower limit on the contribution made by stars to
the Galaxy's central gravitational field.

A concern we had regarding the earlier work by P14 and BP15, was that these
papers assumed that the stars that entered the RAVE survey are not
kinematically biased as a result of some bias in the selection of stars for
spectroscopy. A particular worry was the possibility of bias towards younger
stars, which would be kinematically cooler. A detailed study of the selection
function of the RAVE survey by \cite{Wojno} has now established that the
kinematics of the RAVE stars used in BP15 and here is remarkably
representative of the underlying population.  Specifically, \cite{Wojno} used
the tool {\it Galaxia} \citep{Sharma2011} to form a mock-RAVE sample by
applying the RAVE selection function to a modified Besan\c con Galaxy model
\citep{Robin2003}. Then they compared the kinematics of the selected stars to
those of the underlying population and found that both the giants and the
main-sequence stars had essentially identical kinematics in the mock-RAVE
sample and the underlying population.  \cite{Wojno} find that RAVE's
selection function does introduce a small kinematic bias to the stars in the
turn-off region of the colour-magnitude diagram. In future work we will
extend the present work to dwarf stars, introducing the appropriate bias to
the model before comparing with the data.  We will also update the input data
to RAVE DR5 \citep{Kunder} with the parallaxes and proper motions updated to
the add recently released data from {\it Gaia} \citep{GaiaDR1}.

We have not obtained the posterior probability distribution in model space.
In fact, as Fig.~\ref{fig:logL} suggests, we have probably not even
identified the most probable model. Our searches of model space have been
inadequate because the construction of models specified by DFs $f(\vJ)$ and
have self-consistent gravitational potentials is computationally expensive.
The expense arises from the need to compute the density, by means of a triple
integral over velocity, throughout the huge volume occupied by our Galaxy. As
a result of software improvements since this work commenced, the cost of
model construction has been reduced by about an order of magnitude, and in
future it will be possible to search model space more thoroughly.

\section*{Acknowledgements}

We thank an anonymous referee for comments that enabled us to improve this
paper.
The research leading to these results has received funding from the European Research
Council under the European Union's Seventh Framework Programme (FP7/2007-2013)/ERC
grant agreement no.\ 321067.

Funding for RAVE has been provided by: the Australian Astronomical Observatory;
the Leibniz-Institut f\"ur Astrophysik Potsdam (AIP); the Australian
National University; the Australian Research Council; the French National Research
Agency; the German Research Foundation (SPP 1177 and SFB 881); the
European Research Council (ERC-StG 240271 Galactica); the Istituto Nazionale
di Astrofisica at Padova; The Johns Hopkins University; the National Science
Foundation of the USA (AST-0908326); the W. M. Keck foundation; the Macquarie
University; the Netherlands Research School for Astronomy; the Natural
Sciences and Engineering Research Council of Canada; the Slovenian Research
Agency; the Swiss National Science Foundation; the Science \& Technology
Facilities Council of the UK; Opticon; Strasbourg Observatory; and the
Universities of Groningen, Heidelberg and Sydney. The RAVE web site is at
http://www.rave-survey.org.

\bibliographystyle{mnras}
\bibliography{heatedDM}{}
\end{document}